\documentclass[pre,twocolumn,aps,floatfix,10pt]{revtex4}

\usepackage{amsmath}
\usepackage{amsfonts}
\usepackage{amssymb}
\usepackage{graphicx}
\usepackage{fontenc}
\usepackage{psfrag,layout}
\usepackage{upgreek}

\begin{document}
\title{Crystal-liquid interfacial free energy of hard spheres via a 
novel thermodynamic integration scheme}
 
\author{Ronald Benjamin} % and J{\"u}rgen Horbach}

\email{rbenjamin.phys@gmail.com}
\address{Institut f{\"u}r Theoretische Physik II: Soft Matter,
Heinrich Heine-Universit{\"a}t D{\"u}sseldorf,
Universit\"atsstra\ss e 1, 40225 D{\"u}sseldorf, Germany}

\author{J{\"u}rgen Horbach}
\email{horbach@thphy.uni-duesseldorf.de}
\address{Institut f{\"u}r Theoretische Physik II: Soft Matter,
Heinrich Heine-Universit{\"a}t D{\"u}sseldorf,
Universit\"atsstra\ss e 1, 40225 D{\"u}sseldorf, Germany}

%\affiliation{Institut f{\"u}r Theoretische Physik II: Soft Matter,
%Heinrich Heine-Universit{\"a}t D{\"u}sseldorf,
%Universit\"atsstra\ss e 1, 40225 D{\"u}sseldorf, Germany}
\author{}

\begin{abstract}
The hard sphere crystal-liquid interfacial free energy, ($\gamma_{\rm
cl}$), is determined from molecular dynamics simulations using a novel
thermodynamic integration (TI) scheme. The advantage of this TI scheme
compared to previous methods is to successfully circumvent hysteresis
effects due to the movement of the crystal-liquid interface. This is
accomplished by the use of extremely short-ranged and impenetrable
Gaussian flat walls which prevent the drift of the interface while
imposing a negligible free-energy penalty. We find that it is crucial
to analyze finite-size effects in order to obtain reliable estimates of
$\gamma_{\rm cl}$ in the thermodynamic limit.
\end{abstract}

\maketitle

% footnotes must appear after maketitle
% affiliations first, use letters
%\bgroup
%\renewcommand\thefootnote{\alph{footnote}}
%\footnotetext[1]{%
%Institut f\"ur Theoretische Physik II: Weiche Materie,
%Heinrich Heine-Universit\"at D\"usseldorf, Universit\"atsstra\ss{}e 1,
%40225 D\"usseldorf, Germany}
%\egroup

%\footnotetext[2]{Correspondence: horbach@thphy.uni-duesseldorf.de}

%
\section{Introduction}
\label{sec_intro}
Since the discovery of a fluid-to-solid transition in hard
spheres by computer simulations \cite{alder-wainright57},
the hard sphere model has become one of the
paradigms~\cite{mulerolnp753} for the study of nucleation and crystal
growth~\cite{auerfrenkel2001,lairdaminiprl2006,zykova2009,zykova2010,filion2010}.
The simplicity of the hard sphere interaction potential allows for
the development of theoretical and computational approaches that
allow for quantitative predictions in the context of crystallization
phenomena~\cite{curtin89,marr94,marr95,gruhn2001,snook2003,pusey2009,
tanaka2010,schilling2010}.  The crucial thermodynamic parameter which
governs the mechanism of homogeneous nucleation and subsequent growth of
the crystal from the melt is the crystal-liquid interfacial free energy,
$\gamma_{\rm cl}$, defined as the reversible work required to form an unit
area of a crystal-liquid interface~\cite{adamson97}.  The homogeneous
nucleation rate as well as the final morphology of the resulting
crystal are strongly dependent on the magnitude and anisotropy of this
quantity~\cite{turnbull-cech50,turnbull50,turnbull52,woodruff73,tiller91}.

Several simulation as well as theoretical approaches [based on
Density Functional Theory (DFT)] have attempted to determine the
interfacial free energy of hard-sphere systems, though there have
been some discrepancy in the results obtained from these various
methods~\cite{davidchack-laird2000,cacciuto2003,laird-davidchack2005prl,
laird-davidchack2005jpcb,davmorrislaird2006,davidchack2010,hartel2012,martin2012}.
A direct determination of $\gamma_{\rm cl}$ for hard sphere systems
was made in Ref.~\cite{davidchack-laird2000} using a thermodynamic
integration~\cite{straatsma93,frenkel-smit02} approach known as the
``cleaving wall'' method. Later, the estimates for $\gamma_{\rm cl}$
were revised~\cite{davidchack2010} after fixing an error in the previous
TI scheme.  The same authors carried out TI simulations with the
soft-sphere potential and extrapolated the results to the hard-sphere
limit~\cite{laird-davidchack2005prl,laird-davidchack2005jpcb}. Data in
Ref.~\cite{laird-davidchack2005prl} were a little higher than those
reported for the pure hard sphere system~\cite{davidchack2010}.
The most reliable estimates from an indirect approach based on
capillary fluctuations~\cite{hartel2012} were about $10\%$ higher
than those reported in Ref.~\cite{davidchack2010}. Results from
DFT~\cite{hartel2012}, while in good agreement with the ``cleaving
wall'' and capillary fluctuation methods as regards the anisotropy is
concerned, yielded larger values for the magnitude of $\gamma_{\rm cl}$.
Results from the Tethered Monte Carlo approach are consistent with the
capillary fluctuation method~\cite{hartel2012} but are a few percentages
higher than those reported in Ref.~\cite{davidchack2010}

In the ``cleaving walls"
scheme~\cite{davidchack-laird2000,davidchack2010}, the thermodynamic
integration was carried out by using an external wall, consisting of
particles arranged in an ideal lattice structure, to split the bulk
phases and then join them together.  Finally, the walls were removed
resulting in crystal and liquid phases separated by two interfaces.
In this approach, there were uncontrolled hysteresis errors in the
last step when the external walls were removed.  When both phases are
joined together, the interface is formed at the walls. However, when such
``cleaving walls'' are gradually removed, the interface drifts on account
of thermal fluctuations.  In long simulations, the interfaces can travel
far from the walls by freezing at one end and simultaneously melting at
the other end~\cite{davlaird96,davidchack-laird03,davidchack2010}.  As a
result, the reverse process (when the cleaving walls are reinserted)
does not retrace the same path as the forward process, showing the
existence of hysteresis. This affects the accuracy in the final estimates
of $\gamma_{\rm cl}$.  While the interfacial drift is not a problem for
liquid-liquid interfaces~\cite{schmitz2014}, it is far more severe for
the crystal-liquid interface and needs to be overcome in order to obtain
accurate values for $\gamma_{\rm cl}$.

In a recent work, we have developed a novel TI
scheme to compute $\gamma_{\rm cl}$ for the Lennard-Jones
potential~\cite{benjamin-horbach2014}.  Our method is able to circumvent
problems associated with the drift of the crystal-liquid interface
and provides a better control of hysteresis errors associated with the
latter drift.  The strategy was to use extremely short-ranged and flat
Gaussian walls to constrain the position of the interface while imposing
a negligible free energy penalty.  Another novelty of our scheme was the
use of structured walls consisting of frozen-in crystalline layers to
smoothly transform the system from separate bulk phases to two interfaces
in contact with the bulk fluid and crystal phases.

Apart from the TI scheme, the reliability of $\gamma_{\rm cl}$ estimates
also depends on properly accounting for the finite-size effects. However,
few previous works on the determination of $\gamma_{\rm cl}$ via molecular
simulations include a discussion on finite size effects. In
a recent work, Schmitz {\it et al.}~\cite{schmitz2014}
proposed a scaling relation based on capillary wave theory, to take
into account finite size corrections and get accurate values for the
interfacial free energies in the thermodynamic limit. In our earlier
work~\cite{benjamin-horbach2014} on the crystal-liquid interfacial free
energy for Lennard-Jones systems, we obtained results consistent with
their theory.

In this work, we  compute $\gamma_{\rm cl}$ for hard spheres using
Molecular Dynamics (MD) in combination with TI.  We will determine
$\gamma_{\rm cl}$ for the (100), (110) and (111) orientations of the face
centered cubic crystal-liquid interface.  The hard sphere interactions
are described by a very short-ranged inverse power-law potential (see
below). The reason for using such a short-ranged continuous potential is
the easy adaptability of our TI scheme developed for the continuous LJ
potential into the soft-sphere potential.  Since our TI scheme involves
a direct modification of the interaction potential, it is easier to fit
it into a conventional MD simulation with continuous potential rather
than to a collisional event-driven MD algorithm for a discontinuous
hard-sphere potential~\cite{rapaport04}.

To account for errors due to finite size effects and estimate
$\gamma_{\rm cl}$ in the thermodynamic limit, a careful analysis was
carried out at several system sizes in the framework of capillary wave
theory~\cite{binder82,schmitz2014}. We obtain results consistent with
the predictions of capillary wave theory showing that the introduction
of the flat wall does not suppress capillary fluctuations. The success
of our scheme indicates that our TI approach is also well suited for
very short-ranged potentials.

In the next section, we introduce the potential and then describe the
TI scheme in Sec.~\ref{sec_ti}. The details of the simulation are given
in Sec.~\ref{sec_sim} and results are presented in Sec.~\ref{sec_res}.
Finally, we end with a conclusion in Sec.~\ref{sec_conc}.

\section{Interaction Potential}
\label{sec_model}  
Hard-sphere interactions between a particle $i$ at position $\vec{r}_i$
and a particle $j$ at position $\vec{r}_j$, separated by a distance
$r_{ij} = |\vec{r}_i - \vec{r}_j|$ are approximated by the inverse
power-law potential
\begin{equation}
\phi(r_{ij}) = 
\epsilon \left(\frac{\sigma}{r_{ij}}\right)^{n} 
\quad \quad {\rm with} \quad \quad n=256,
\label{eq:inv_power_pot}
\end{equation}
where $\epsilon$ and $\sigma$ set the energy and length scale,
respectively.  For computational efficiency, the potential was cut-off
at a distance $r_{\rm c}=1.2\,\sigma$, where the potential has a
value $10^{-21} \epsilon/k_{\rm B}T$.  With the exponent $n=256$, the
parameters for solid-fluid coexistence at the temperature $k_{\rm B} T=
1.0\,\epsilon$ (with $k_{\rm B}$ the Boltzmann constant) are very close
to those for the hard-sphere system~\cite{zykova2010}, in agreement with
recent findings \cite{lange2009} (see below). 

\section{Method}
\label{sec_ti}
The interfacial free energy $\gamma_{\rm cl}$ is the excess free-energy
per area that results from the formation of an interface between the
crystal and the liquid phase.  It can be expressed via the difference
between the free energy of the inhomogeneous system with crystal-liquid
interface, $F_{\rm cl}$, and the sum of the bulk free energies of the
crystal and liquid, $F_{\rm c}$ and $F_{\rm l}$, respectively,
\begin{equation}
\gamma_{\rm cl}=\frac{F_{\rm cl}-(F_{\rm c}+F_{\rm l})}{A}
\end{equation}
with $A$ the area of the interface.

Computing $\gamma_{\rm cl}$ via thermodynamic integration involves
joining together bulk crystal and liquid phases at coexistence to form
an inhomogeneous system involving the individual phases separated by
two interfaces. To ensure a path with minimal hysteresis, the crystal
phase should be perturbed as little as possible such that no stress is
generated in the crystal when it comes into contact with the liquid phase.

\begin{figure}
\includegraphics[width=3.0in,scale=0.5]{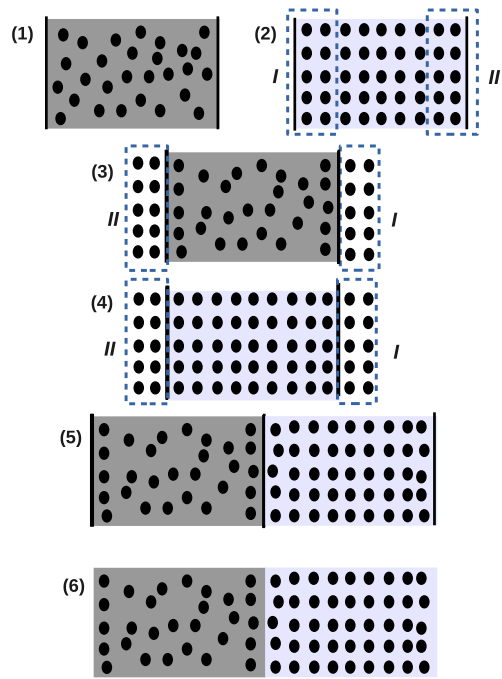}
\caption{\label{fig:TIscheme}
Schematic of the TI scheme adopted in this work.  Particles in the light
and dark simulation boxes represent the crystal and the liquid phases,
respectively. For details see text.}
\end{figure}
Here, we provide a novel TI scheme to compute $\gamma_{\rm cl}$
for the inverse power potential, Eq.~(\ref{eq:inv_power_pot}). Our
TI scheme is based on an earlier approach used to obtain
the crystal-liquid interfacial free energy for a Lennard-Jones
potential~\cite{benjamin-horbach2014}. Compared to previous TI methods
such as the ``cleaving potential"~\cite{broughton-gilmer86} or ``cleaving
walls"~\cite{davidchack-laird2000,davidchack-laird03,laird-davidchack2005prl,laird-davidchack2005jpcb,davidchack2010}
scheme, our approach differs in two crucial ways.  Firstly, in order
to achieve minimal perturbation of the crystalline phase during the
transformation, the liquid should be ordered into a crystal-like layers
near the boundaries of the simulation box and this ordering must be
compatible with the actual structure of the crystal phase.  In our
TI scheme, this is achieved by introducing a structured solid wall,
consisting of particles belonging to the crystal phase, frozen into their
instantaneous equilibrium position. Interactions between the particles
in the bulk phases and the structured wall are of the same kind as that
between the bulk particles.

Secondly, to tackle the problem associated with the movement of the
crystal-fluid interface due to thermal fluctuations~\cite{davlaird96},
a Gaussian flat wall is introduced to prevent the interface from
drifting during the simulation.  However, unlike the previous TI scheme
\cite{benjamin-horbach2014}, where an extremely short-ranged flat wall
is introduced at the beginning of the scheme, here in the first step we
introduce a flat wall with a similar range as the interaction potential
between the particles.  Only in the final step, the range of this wall
is gradually reduced to a value much smaller than the effective size
of the particles $\sigma$ (about $10^4$ times less).  This trick saves
computational time as a short-time step needs to be used in the final step
to integrate the short-ranged forces. The use of a multiple time-step
algorithm~\cite{frenkel-smit02} in the final step further improves the
computational efficiency of the scheme.

Initially, bulk liquid and crystal phases are simulated in a box
with dimensions $L_{\rm x}\times L_{\rm y}\times L_{\rm z}$ such that
the two phases have the same volume but different particle numbers.
The final state consists of two crystal-liquid interfaces connecting the
bulk phases, with a total length $2L_{\rm z}$ along the $z$ direction.
The TI scheme proceeds via six steps and in each step the transformations
are carried out by directly modifying the interaction potential by a
parameter $\lambda$.  This idea is similar to the TI scheme presented
in earlier works to compute $\gamma_{\rm cl}$ and the interfacial free
energies of liquid and crystal phases in contact with flat and structured
walls~\cite{benjamin-horbach2012,benjamin-horbach2013,benjamin-horbach2014}.
The specific choices of the $\lambda$ parameterizations adopted below,
yielded smooth thermodynamic integrands allowing for an accurate numerical
calculation of the integrals. In the following, we outline the TI scheme,
in detail.

{\it{Steps 1 and 2}}: The initial thermodynamic state of our system
consists of separate bulk liquid and crystal phases with periodic boundary
conditions (along the $z$ direction), while the final state involves
liquid and crystal separated by two interfaces. To reach the final state,
at some point during the transformation, interactions between the two
sides of each phase through the periodic boundaries must be switched
off while the interactions between the two phases must be turned on.
While rearranging the periodic boundaries, it must be ensured that
the particles belonging to each phase remain inside their respective
simulation cells and do not cross the boundaries such that density of
each phase in its box remains at the respective coexistence density.
For this purpose, in the first and second steps, a Gaussian flat
wall is gradually introduced at both ends of the liquid and crystal
simulation boxes respectively, along the $z$ direction [sketch (1)
in Fig.~\ref{fig:TIscheme}]. The periodic boundary conditions are kept
intact.  This  wall acts as a barrier to prevent the liquid and crystal
particles from crossing the boundaries.  The transformation is carried
out by gradually increasing the height of the potential barrier.

The interaction of a particle $i$ with the flat wall is modeled by a 
Gaussian potential,
\begin{equation}
\label{eq:flatgwall}
u_{\text{fw}}(z_{iw})=
a \exp\left[-\left(\frac{z_{iw}}{b}\right)^{2}\right] \, ,
\end{equation}
with $z_{iw}$ the distance of the particle from the wall in $z$-direction.
The height and range of the Gaussian potential is determined by the
variables $a$ and $b$. At a temperature $T=1.0$, we chose $a=25 k_{\rm
B}T/\epsilon$, such that the particles cannot overcome the barrier. $b$
was set to $0.027\,\sigma$. For this value of $b$, the bulk density does not
change in the presence of the wall.

The parameter $\lambda$ is coupled to the flat wall as follows,
\begin{equation}
\begin{split}
\label{eq:uwallsubst}
u_{\text{fw}}(\lambda,z_{iw})=& \lambda^{2} u_{\text{fw}}(z_{iw})
\end{split}
\end{equation}
At $\lambda=0$, there is no wall and as $\lambda$ increases the wall
becomes more and more impenetrable.

The $\lambda-$dependent Hamiltonian for steps 1 and 2 takes the form
\begin{equation}
\begin{split}
H_{1(2)}(\lambda) = \sum_{i=1}^{N}
\frac{{{\bf p}_{i}^{2}}}{2m}
+ {U_{\text{pp}}^{\text{l(c)}}} 
+ \lambda^{2}{U_{\text{fw}}^{\text{l(c)}}}\, ,
\end{split}
\label{eq:hamilt_step1}
\end{equation}
where, $H_{\rm 1(2)}$ represents the Hamiltonian for interaction of
the flat wall with the liquid (crystal) and ${\bf p}_{i}$ and $m$
denote the momentum and mass of particle $i$ with all particles having
the same mass. The potential energy due to particle-particle interactions is represented by
${U_{\text{pp}}^{\text{l(c)}}}= \sum_{i=1}^{N^{\text{l(c)}}} \sum_{j=i+1}^{N^{
\text{l(c)}}}u(r_{ij})$.  The interaction
potential of particles with the flat wall is denoted by
${U_{\text{fw}}^{\text{l(c)}}}=\sum_{i=1}^{N^{\text{l(c)}}}u_{\text{fw}}(z_{
\text{iw}})$, where the superscript l(c) refers to particles in liquid
(crystal) phase and $N^\text{l(c)}$ is the total number of liquid
(crystal) particles.

{\it Steps 3 and 4}: From the two sides of the crystal simulation
cell, containing $1-2$ layers of crystalline particles frozen into an
instantaneous equilibrium configuration, two structured walls were
constructed.  As shown in Fig.~\ref{fig:TIscheme}, these walls are
juxtaposed at the appropriate ends of the liquid and crystal simulation
cells corresponding to steps 3 and 4, respectively. The flat walls
are still present to prevent particles from crossing the boundaries.
During the transformation, the structured walls are gradually switched
on, while interactions through the periodic boundaries along the $z$
direction are gradually switched off.

The $\lambda$-dependent Hamiltonian for step 3 is given by,
\begin{equation}
\begin{split}
H_{3}(\lambda) =  
\sum_{\text{i=1}}^{N} \frac{{\bf p}_i^{2}}{2m}
+ {{U}_{\text{pp}}^{\text{l}}}
+ (1-\lambda)^{256}{{U^{\ast}}_{\text{pp}}^{\text{l}}}\\
+ \lambda^{256}{U_{\text{pw}}^{\text{l}}}
+ {U_{\text{fw}}^{\text{l}}}
\end{split}
\label{eq:hamilt_step3}
\end{equation}
Since the inverse-power potential (Eq.~\ref{eq:inv_power_pot}) is
very short-ranged, and the liquid particles being present close to
the boundaries of the simulation box at the beginning of this step,
a rapidly decaying $\lambda$-function for the periodic boundaries
ensures a smoothly varying thermodynamic integrand. Effectively,
this transformation is carried out by gradually modifying the size
of the particles, i.e.~$\epsilon([(1-\lambda) \sigma]/r)^{256}$ and
$\epsilon([\lambda \sigma]/r)^{256}$ for switching off the periodic
boundaries and switching on the structured walls, respectively.

In the ``cleaving walls" TI scheme, the interaction between the individual
phases and the walls was brought about by moving the walls towards
the bulk liquid and crystal phases. This required the use of a complex
``corrugated cleaving plane"~\cite{davidchack-laird03,davidchack2010},
which is compatible with the structure of the wall. However, in our TI
scheme, the interaction between the two phases or between each individual
phase and the structured walls takes place across a flat plane. As
shown above, this is achieved by directly modifying the interaction
potentials to carry out the transformations, leading to a simple and
straightforward algorithm.

The corresponding Hamiltonian for step 4 is given by,
\begin{equation}
\begin{split}
H_{4}(\lambda) =  
\sum_{\text{i=1}}^{N} \frac{{\bf p}_i^{2}}{2m}
+ {{U}_{\text{pp}}^{\text{c}}}
+ (1-\lambda)^{8}{{U^{\ast}}_{\text{pp}}^{\text{c}}}\\
+ \lambda^{256}{U_{\text{pw}}^{\text{c}}}
+ {U_{\text{fw}}^{\text{c}}}
\end{split}
\label{eq:hamilt_step4}
\end{equation}
Since it is necessary to maintain the crystalline structure throughout
the transformation, in Eq.~(\ref{eq:hamilt_step4}), the periodic
boundaries are swit\-ched off gradually such that these interactions become
weaker only when the structured walls are already strongly
interacting with the crystal phase.

In Eqs.~(\ref{eq:hamilt_step3}) and (\ref{eq:hamilt_step4}),
${{U^{\ast}}_{\text{pp}}^{\text{c(l)}}}$ specifies the periodic boundary
interactions, while ${U_{\text{pp}}^{\text{c(l)}}}$ corresponds to
the bulk interactions.  Interactions between the individual phases
and the structured wall particles is denoted by $U_{\text{pw}}$,
where ${U_{\text{pw}}^{\text{c(l)}}}=\sum_{i=1}^{N^{\text{c(l)}}}
\sum_{j=1}^{N^{\text{w}}}u_{\text{pw}}(r_{{ij}})$, with $N^{\text{w}}$
being the total number of particles in the structured walls. The same
inverse power potential was used for $u_{\text{pw}}$, as given by
Eq.~(\ref{eq:inv_power_pot}), with the parameter $\epsilon$ replaced by
$\epsilon_{\rm pw}$.  Throughout the transformation in steps 3 and 4 as
well as in subsequent steps, $\epsilon_{\rm pw}/\epsilon=1$ was kept
constant, where $\epsilon$ refers to the interaction strength between
the particles of the system.

{\it Step 5}: In this step, the individual liquid and crystal phases
are gradually brought together in the presence of the Gaussian flat
walls, while the structured walls are removed.  Since the periodic
boundary conditions in the two phases are already switched off, only
interactions between the two phases need to be turned on [see (5)
in Fig.~\ref{fig:TIscheme}]. At the end of this step, the resulting
thermodynamic state consists of bulk crystal and liquid phases separated
by two interfaces whose position is tied to the position of the flat
walls.

The Hamiltonian corresponding to step 5 is
\begin{equation}
\begin{split}
H_{5}(\lambda) =  
\sum_{\text{i=1}}^{N_{p}} 
\frac{{\bf p}_{\text{i}}^{2}}{2m_{i}}
+ {{{U}_{\text{pp}}^{\text{c(l)}}}}
+ \lambda^{2} \left(\frac{1+\lambda}{2}\right)^{\rm 256}   
  {U_{\text{pp}}^{\text{c+l}}}\\
+ (1-\lambda)^{2} (1-\lambda/2)^{256}  
  {U_{\text{pw}}^{\text{c(l)}}}
+ {U_{\text{fw}}^{\text{c(l)}}},
\end{split}
\label{eq:hamilt_step5}
\end{equation}
where the interaction potential between the liquid and crystal phases
is denoted by, ${U_{\text{pp}}^{c+l}}= \sum_{i=1}^{N^{\text{l}}}
\sum_{j=1}^{N^{\text{c}}}u(r_{{ij}})$. In Eq.~(\ref{eq:hamilt_step5}),
there is a one-to-one correspondence in the $\lambda$-parameterizations
such that interactions between the crystal and liquid phases are turned
on while interactions with the structured wall are switched off to ensure
a smoothly varying thermodynamic integrand.

{\it Step 6}: The last step involves removing the flat walls and it
is difficult to achieve total control over the reversibility of the
scheme. Due to thermal fluctuations, the two interfaces can move by
melting on one side and refreezing on the other side, if the potential
barrier due to the walls is weak enough, i.e.~($a \leq k_{\rm B}T$).
However, this makes the transformation irreversible since if the walls
are reinserted, the position of the interfaces will not coincide with the
position of the walls. As a result, in the two simulation boxes a mixture
of liquid and crystal phases will be obtained and one cannot retrace
the previous steps in the reverse direction to reach the initial state
consisting of independent liquid and crystal phases at their respective
coexistence densities.

However, by a clever scheme one can reduce the error due to the resulting
hysteresis in the TI path to a negligible value such that it does
not affect the accuracy in the final estimates of $\gamma_{\rm cl}$.
This is achieved by breaking step 6 into two sub-steps 6a and 6b.
In the first sub-step, the range of the Gaussian flat wall is gradually
reduced while maintaining the same height for the potential barrier. In
principle, this sub-step is reversible since, the particles still cannot
cross the flat wall barrier at the end of this step.  In the next step,
6b, the height of the barrier is slowly switched off such that in the
end one has the desired state consisting of liquid and crystalline
phases separated by two interfaces.  This transformation is no more
reversible due to the movement of the interface.  However, if the
range of the Gaussian flat wall has been reduced significantly in
step 6a such that very few particles interact with the wall,
the contribution of step 6b would be negligible in comparison to
the magnitude of $\gamma_{\rm cl}$ and can be ignored.

The Hamiltonian corresponding to sub-step 6a is given by
\begin{equation}
\begin{split}
H_{6a}(\lambda) = \sum_{i=1}^{N}
\frac{{{\bf p}_{i}^{2}}}{2m}
+ {U_{\text{pp}}^{\text{l(c)}}} 
+ {U_{\text{fw}}^{\text{l(c)}}}(\lambda)\, ,
\end{split}
\label{eq:hamilt_step6a}
\end{equation}
with
\begin{equation}
\begin{split}
\label{eq:uwallfwstep6a}
u_{\text{fw}}(z_{iw},\lambda)=&
a\exp(-[z_{iw}/b(\lambda)]^{2}) \, .
\end{split}
\end{equation}
where ${U_{\text{fw}}^{\text{l(c)}}}(\lambda)=\sum_{i=1}^{N_{\rm l(c)}}
u_{\text{fw}}(z_{iw}, \lambda)$.  In Eq.~(\ref{eq:uwallfwstep6a}),
$b(\lambda) = {(1-\lambda)}^{2}b^{\prime}$ with $b^{\prime}=0.027\sigma$
and $\lambda$ varies from $0$ to $0.893$ such that the parameter
$b(\lambda)$ is reduced from $0.027\sigma$ to $0.0003\sigma$.

In the next sub-step, 6b, the extremely short-ranged Gaussian walls are
gradually switched off. The Hamiltonian for this sub-step is
\begin{equation}
\begin{split}
H_{6b}(\lambda) = \sum_{i=1}^{N}
\frac{{{\bf p}_{i}^{2}}}{2m}
+ {U_{\text{pp}}^{\text{l(c)}}} 
+ {(1-\lambda)^2} {U_{\text{fw}}^{\text{l(c)}}}\, ,
\end{split}
\label{eq:hamilt_step6b}
\end{equation}
In this last step, $b$ was kept at the same value as at the end of step
6a, viz.~$b=0.0003\sigma$.  In principle, the flat wall could be made even more
short-ranged by varying $\lambda$ in step 6a up to, say $0.99$. However,
our simulations showed that reducing the range of the flat wall to such
a low value is unnecessary for our purpose, since the combined numerical
and statistical errors from steps one to five would be much larger than
the total contribution from step 6b.

Generally, the contribution of step 6b, depends on the average density
of particles in the interface region. Prior work on the crystal-liquid
interface corresponding to hard sphere systems has shown that density
near the interface is the mean of the bulk liquid and solid coexistence
densities~\cite{davlaird96}.  Even accounting for capillary fluctuations,
the average density will not be very far from this mean value. Therefore,
the contribution of step 6b will be close to that obtained for the flat
wall-liquid interface.  For example, at $b=0.0003 \sigma$, the excess
free-energy of the fluid in contact with the flat-wall was about $10$
times less than the statistical errors from steps 1-6a. For the crystal,
the excess free energy was about $1000$ time less than the typical
statistical errors.  Since our simulations yield such a small free energy
difference for liquid in contact with the flat wall at $b=0.0003\sigma$,
it is clear that step 6b will yield a similar negligible value and
hence this step can effectively be ignored. In general, our simulations
indicate a negligible value (less than $10^{-4} k_{\rm B}T$) for the
liquid-flat wall free energy difference and hence that of step 6b,
as well, if $b\sim 10^{-4}\sigma$.

We carried out several independent runs for one system size, both in the
forward as well as reverse directions to check the reversibility of step
6b and obtained the free-energy difference from runs which yielded the
least hysteresis \cite{davidchack2010,davidchack-laird03}. Our simulation
results showed a negligible contribution for this step of the order of
$0.00005$ $k_{\rm B}T$ for the (100) and (111) orientations and of the
order $0.0005k_{\rm B}T$ for the (110) orientation. Since, the combined
statistical and numerical errors in the previous steps (1-6a) are in
the range $0.003-0.005 k_{\rm B}T$, this last step was not performed
for other system sizes at which simulations were carried out.

The free energy difference in the various steps can be computed as
\begin{equation}
\Delta F_{\rm i} = 
\int_{0}^{1} \left\langle \frac{\partial H_{\rm i}}{\partial \lambda} 
\right\rangle d\lambda
\end{equation}
where $i$ varies from $1$ to $6$ corresponding to the six steps.

The interfacial free energy $\gamma_{\rm cl}$ is obtained by adding the
the free energy differences corresponding to the six steps, divided by
the total interfacial area $A$,
\begin{equation}
\gamma_{\rm cl} = \frac{ \sum_{i=1}^6 {\Delta F}_{\rm i}}{A} \, ,
\end{equation}
with $A=2L_{\rm x}L_{\rm y}$ (the factor 2 takes into account
the presence two independent planar crystal-liquid interfaces,
cf.~Fig.~\ref{fig:TIscheme}).

\section{Simulation Details}
\label{sec_sim}
Molecular Dynamics (MD) simulations were carried out in the canonical
ensemble with the total number of particles $N=N_{\rm l}+N_{\rm c}$,
volume $V$ and temperature $T$ maintained constant. Constant temperature
was achieved by assigning every 200 time steps random velocities to the
particles distributed according to the Maxwell-Boltzmann distribution.
Newton's equations of motion were integrated according to the velocity
Verlet algorithm~\cite{allen-tildesley87}. During steps one to five
of our TI scheme, a time-step $\Delta t_{\rm large}=0.0005 \tau$ (with
$\tau=\sqrt{(m{\sigma}^{2}/\epsilon)}$) was used. In the
sixth step, to take into account the extremely short-range forces due to
the Gaussian flat wall, a multiple time-step scheme~\cite{frenkel-smit02}
was applied, where in combination with $\Delta t_{\rm large}$, a smaller
time-step of $\Delta t_{\rm small}=0.000025\,\tau$ was used.  It was
observed that this slowed down the simulations by approximately a factor
of two.

The coexistence densities for the inverse-power potential,
Eq.~\ref{eq:inv_power_pot}, as computed using the free-solidification
method~\cite{zykova2010,kuhn13}, at the temperature $T = 1.0$, are
$\rho_{\rm l}^{\rm inv}=0.933\,\sigma^{-3}$ for the liquid phase and
$\rho_{\rm c}^{\rm inv}=1.030\,\sigma^{-3}$ for the crystal phase; the
coexistence pressure is given by $P_{\rm co}^{\rm inv}=11.46 k_{\rm
B}T/{\sigma}^{3}$~\cite{undulating_subst_2014}.  In comparison,
the coexistence parameters of the actual hard-sphere system are
$\rho_{\rm l}^{\rm HS}=0.940 {\sigma}^{-3}$, $\rho_{\rm c}^{\rm
HS}=1.041{\sigma}^{-3}$ and $P_{\rm co}^{\rm HS}=11.576 k_{\rm
B}T/{\sigma}^{3}$~\cite{zykova2010}.

From these co-existence values one can define an effective dimensionless
diameter $\sigma^{\rm eff}=\sigma (\rho_{\rm c}^{\rm HS}/\rho_{\rm c}^{\rm
inv})^{-1/3}$, which can be used as a scaling parameter to compare the
two systems. Such an effective diameter can also be obtained using
the co-existence pressure value as well as the coexistence density
of the liquid.  It is observed that using the crystal co-existence
density and the co-existence pressure as the scaling variable leads to
the same effective diameter, viz.~$\sigma^{\rm eff}= 1.0035\sigma$,
while use of the liquid co-existence density leads to a slightly lower
effective diameter $\sigma^{\rm eff}= 1.0024\sigma$.  We will use the
effective diameter value $\sigma^{\rm eff}= 1.0035\sigma$ to compare
the crystal-liquid interfacial free energy obtained for the inverse
power-law model with the hard-sphere values. The equivalent values for the
hard-sphere model are obtained as $\gamma_{\rm cl}^{\rm HS}=\gamma_{\rm
cl}^{\rm inv} {{\sigma^{\rm eff}}^2}/{k_{\rm B}T}$, where $\gamma_{\rm
cl}^{\rm inv}$ is the interfacial free energy corresponding to the
inverse power-law potential.

We obtain $\gamma_{\rm cl}$ for the (100), (110) and (111) orientations
of the face-centered cubic crystal with respect to the liquid at the
interface. To study finite-size effects, simulations were carried out at
various system sizes for the (100) orientation of the interface, ranging
from around $7000$ to $34000$ particles, with various lateral dimensions and
with a total longitudinal dimension of about $80\sigma$. For the other
two orientations, simulations were carried out only at the largest
system sizes. The dimensions of our system along with the total number
of particles are specified in Table I.

To generate initial configurations, liquid and crystal phases were
equilibrated for about $2\times 10^6$ time steps (in steps of ${\Delta
t}_{\rm large}$) at their respective co-existence densities. Both phases
were simulated in cells of identical dimensions with $N_{\rm l}<N_{\rm
c}$, since $\rho_{\rm l}<\rho_{\rm c}$.  To calculate $\gamma_{\rm cl}$
via TI, independent runs were carried out at $50-100$ values equally
spaced intervals of $\lambda$ between the initial and final states
in order to obtain smooth thermodynamic integrands. At every value of
$\lambda$, the system was first equilibrated at $\lambda=0$ for $10^4$
time steps and then $\lambda$ was continuously increased until the desired
value of $\lambda$ was reached. The number of time steps to carry out
this switch varied from $2.5\times10^5$ to $10^6$ time steps for the
various TI steps. After the final value $\lambda_{i}$ was reached,
the system was further equilibrated for times varying from $10^6$ to
$4\times 10^6$ time steps. Then the production runs were carried out
over a period varying from $10^6$ to $5\times 10^6$ time-steps for the
different TI steps to obtain the desired statistical accuracy.

For steps 3, 4 and 5, a cubic spline interpolation of the bare data was
performed to obtain the thermodynamic integrand at 100 intervals between
$\lambda=0$ and $\lambda=1$.  Then, Simpson's rule was used to numerically
calculate the integral.  For steps 1, 2 and 6, the thermodynamic integrals
were calculated numerically using the trapezoidal rule from the bare data.
Statistical errors were calculated by partitioning the production runs
into $5$ blocks and then determining the standard deviation from these
$5$ samples.

To check the reversibility of each step in the TI scheme both simulations
were also carried out in the reverse direction, to detect any hysteresis
in the transformation. The initial state for the reverse TI simulations
corresponded to the final state of the forward TI path.  The final values
for $\gamma_{\rm cl}$ reported in Table I correspond to an average of
the free energy differences obtained from the forward and reverse TI
simulations.

\section{Results}
\label{sec_res}
\begin{figure}
\includegraphics[width=3.0in,scale=0.5]{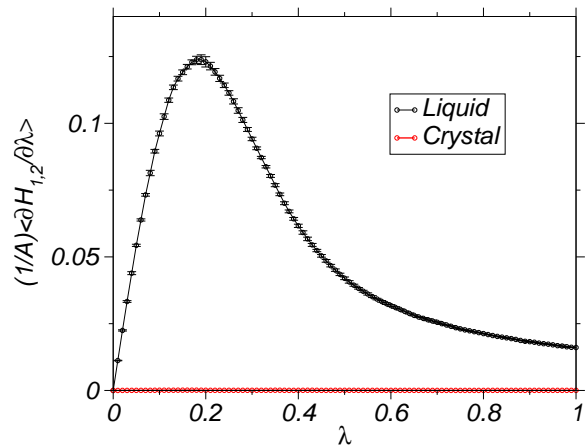}
\caption{\label{fig:step12TI}
Thermodynamic integrand for steps 1 and 2 for the liquid and crystal
phases, respectively, at $T=1.0$ and the (100) orientation of the
crystal. Error bars in this figure and subsequent ones represent one
standard deviation.}
\end{figure}

\begin{figure}
\includegraphics[width=3.2in]{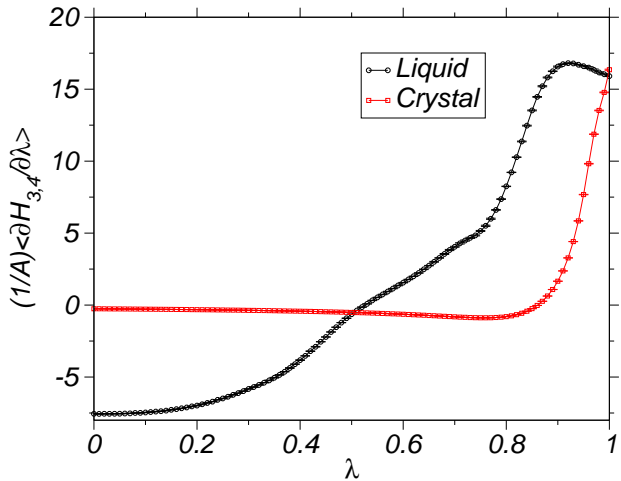}
\caption{\label{fig:step34TI}
Thermodynamic integrand corresponding to steps 3 and 4, for the liquid
and (100) orientation of the crystal, respectively, at the temperature
$T=1.0$.}
\end{figure}

\begin{figure}
\includegraphics[width=3.0in]{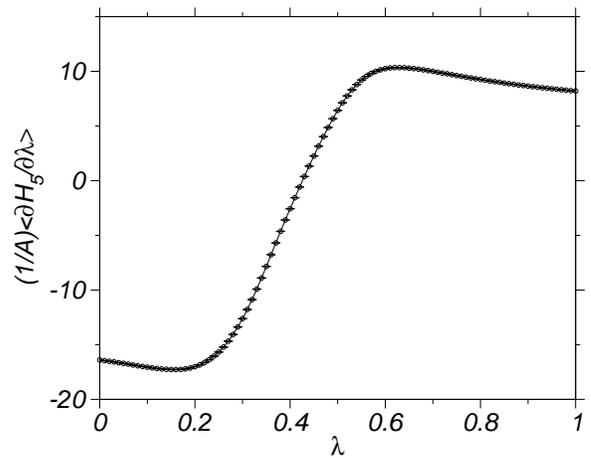}
\caption{\label{fig:step5ti}
Thermodynamic integrand for step 5 at $T=1.0$, bringing the (100)
orientation of the crystal in contact with the liquid.}
\end{figure}

\begin{figure}
\includegraphics[width=3.0in]{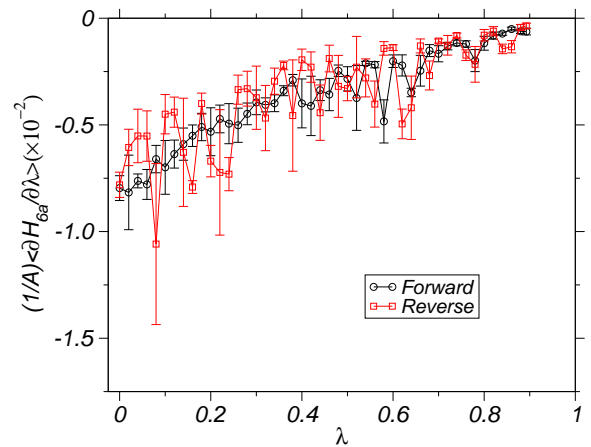}
\caption{\label{fig:step6TI}
Thermodynamic integrands for the forward and reverse processes
corresponding to step 6a, towards reducing the range of the flat wall.}
\end{figure}

The thermodynamic integrands for the various steps are plotted in
Figs.\ref{fig:step12TI} to \ref{fig:step6TI}, for the (100) orientation
of the crystal-liquid interface. The data corresponds to the largest
system size of $32205$ particles with dimensions $20.43 \times 20.43
\times 78.58$ (in units of $\sigma^{-3}$).  Figure~\ref{fig:step12TI}
shows the thermodynamic integrand corresponding to steps 1 and 2 of our
scheme. It is clear that the area under the thermodynamic integrand curve
corresponding to the crystal is negligible as compared to the liquid.
Since the crystal is positioned symmetrically with respect to the two
ends of the simulation cell, the location of the flat walls (at $z=0$
and $L_{\rm z}$) coincides with a density minimum between two crystalline
layers.  Moreover, the crystal has a small diffusivity as compared to
the liquid. Therefore, the free energy cost of inserting a flat wall in
the crystal is negligible compared to that in the liquid.

For the (100), (110) and (111) orientations of the crystal, ${{\Delta
F_{2}}/A}=0.00004 \pm 0.00001$, $0.0018 \pm 0.00004$ and $7.66 \times
10^{-7} \pm 1.94 \times 10^{-7}$ while ${{\Delta F_{1}}/A}=0.053
\pm 0.0002$ (in units of $k_{{\rm B}}T/\sigma^{2}$).  The above data
corresponded to $b=0.027\sigma$.  The smoothness of the integrand and the
smallness of the error bars in Fig.~\ref{fig:step12TI} indicates the lack
of hysteresis in our TI scheme.  Due to the perfect overlap between the
forward and reverse thermodynamic integrands, Fig.~\ref{fig:step12TI}
shows only the curves corresponding to the forward transformation.
For step 1, the difference between the forward and reverse TI results was
$0.0002 k_{\rm B}T/\sigma^2$, and for the different orientations in step
2, the hysteresis was always less than $0.00004 k_{\rm B}T/\sigma^{2}$.

It is to be noted that unlike in our previous
work~\cite{benjamin-horbach2014}, where an extremely short-ranged wall
was inserted from the very first step onwards, yielding a negligible
contribution ($\Delta F_{\rm 1}/A$ was less than $0.002 k_{\rm
B}T/\sigma^{2}$), here step 1 yields a free-energy difference which cannot
be ignored (almost $10\%$ of the final value for $\gamma_{\rm cl}$). This
is because the Gaussian flat wall has a slightly longer range (of the
same order as the interaction potential itself).  The reason for inserting
this slightly longer ranged wall, apart from computational efficiency as
mentioned previously, has to do with the hysteresis observed in step 3,
when the liquid is arranged into an ordered structure near the interface.
A relatively longer-ranged flat wall induces, at the end of step 1,
layering in the liquid near the flat wall, which is compatible with
the ordered structure that would be formed near the interface by the
structured wall. This results in a smoother thermodynamic integrand.
However, in presence of an extremely short-ranged wall (say, for example
with $b= 0.0003\,\sigma$), the structure near the interface, at
the end of step 1, is the same as in the bulk.  This leads to hysteresis
errors in step 3 and the transformation is not smooth anymore, unless
the system is equilibrated for a sufficiently long time.

In Fig.~\ref{fig:step34TI}, thermodynamic integrands corresponding to
steps 3 and 4 are shown.  We obtained excellent overlap between the
forward and reverse thermodynamic integrand curves and therefore only
data corresponding to the forward transformation is reported.  For step
3, hysteresis between the forward and reverse TI calculations was less
than $0.005 k_{\rm B}T$, while for step 4 the same was less than $0.001
k_{\rm B}T$ for all the three orientations.

At the end of step 3, the liquid is ordered into crystalline
layers near the interface. Therefore, the structured wall induces
precrystallization of the liquid even at the bulk liquid coexistence
density~\cite{heniloewenjpcm}.  Since the liquid is already ordered
into a structure compatible with the crystal, the next transformation,
step 5, corresponding to joining the liquid and crystal phases, occurs
smoothly as shown in Fig.~\ref{fig:step5ti}.  The hysteresis between
the forward and reverse TI paths was less than $0.007 {k_{\rm B}}T$
for the various orientations and system-sizes specified in Table I.

The final step consists of two sub-steps. In step 6a, we reduce the range
of the Gaussian flat wall to an extremely small value. As specified
in Section III, the parameter $b$ was changed from $0.027\sigma$
[corresponding to a range of $0.1\sigma$ with the range considered to
be the distance at which $u_{fw}(z_{\rm iw})$ decays to about $0.0001
k_{\rm B}T$] to $0.0003\sigma$ (corresponding to a range of about $0.001
\sigma$).  For this range of the potential, the free-energy difference
per unit area of a liquid in contact with this flat wall was $0.0005k_{\rm
B}T$ and for the various orientations of the crystal is less than $0.00002
k_{\rm B}T$. Clearly, reducing the range of the flat wall by about $100$
times also reduces the free-energy difference by about the same factor.

\begin{figure}
\includegraphics[width=3.0in]{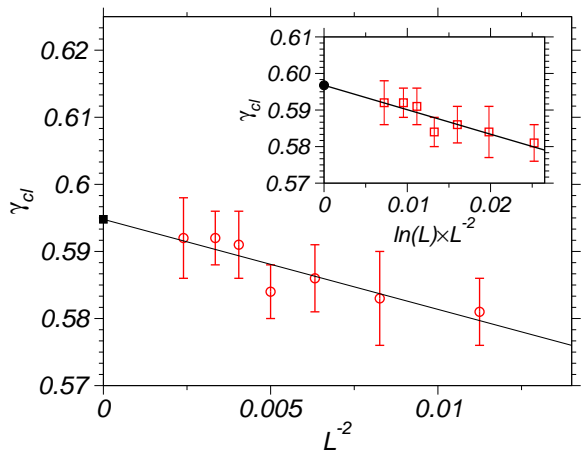}
\caption{\label{fig:gam_vs_l2inv}
$\gamma_{\rm cl}$ as a function $1/L^{2}$, where $L=L_{\rm x}=L_{\rm
y}$, for (100) orientation of the crystal-liquid interface with
the longitudinal dimension $2L_{\rm z}=78.58\,\sigma$. Inset shows
$\gamma_{\rm cl}$ as a function of ~$\ln(L)/L^{2}$. Solid symbols
correspond to the $\gamma_{\rm cl}$ values in the thermodynamic limit.}
\end{figure}
\begin{table*}[htbp]
\centering
\begin{tabular}{|c|c|c|c|c|c|c|}
\hline
\cline{1-5}
 $\text{Orientation}$&$\text{N}$&$\text{System Size}$&$\gamma_{\rm cl}^{\rm inv}$&$\gamma_{\rm cl}^{\rm HS}$ &Cleaving (HS)&Cleaving (IP)\\
\hline \hline
$100$&$6860$&$9.43 \times 9.43 \times 78.58$&$0.581\pm 0.005$&$0.585\pm 0.005$&---&---\\
$100$&$9338$&$11.0 \times 11.0 \times 78.58$&$0.584\pm 0.007$&$0.588 \pm 0.007 $&---&---\\
$100$&$12196$&$12.57 \times 12.57 \times 78.58$&$0.586\pm 0.005$&$0.590 \pm 0.005$&---&---\\
$100$&$15436$&$14.15 \times 14.15 \times 78.58$&$0.584\pm 0.004$&$0.588 \pm 0.004 $&---&---\\
$100$&$19056$&$15.72 \times 15.72 \times 78.58$&$0.591\pm 0.005$&$0.595 \pm 0.005$&---&---\\
$100$&$23058$&$17.29 \times 17.29 \times 78.58$&$0.592\pm0.004$&$0.596\pm 0.004 $&---&---\\
$100$&$32205$&$20.43 \times 20.43 \times 78.58$&$0.592\pm0.006$& $0.596 \pm 0.006$&$0.5820\pm0.0019 $& $0.592\pm 0.007 $\\
$100$&$\infty $&$ \infty $&$0.595^{\rm a},0.597^{\rm b}$&$0.599^{\rm a},0.601^{\rm b}$&---&---\\
$110$&$32106$&$20.43 \times 20.0 \times 80.02$&$0.573\pm 0.004$&$0.577\pm0.004$&$0.5590\pm 0.0020 $&$0.571\pm 0.006$\\
$111$&$33959$&$21.17 \times 20.0 \times 81.66$&$0.552\pm 0.003$&$0.556\pm0.003$&$0.5416\pm 0.0031 $&$0.557\pm 0.007$\\
\hline
\end{tabular}
\caption{Interfacial free energy $\gamma_{\rm cl}^{\rm inv}$ for different
system sizes corresponding to various orientations of the crystal-liquid
interface. The interfacial free-energies for the pure hard-sphere system
obtained by using a scaling parameter are also shown (see text).  For
comparison, data from the cleaving wall methods for the pure hard-sphere
(HS) system~\cite{davidchack2010} and for the inverse power-law (IP)
potential in the hard-sphere limit~\cite{laird-davidchack2005prl} are also
reported.  Dimensions ($L_{\rm x } \times L_{\rm y} \times L_{\rm z}$) are
in units of $\sigma^{3}$. For the (100) orientation of the crystal-liquid
interface, the interfacial free energy in the thermodynamic limit is
extrapolated from the  values of $\gamma_{\rm cl}$ at the various system
sizes using the (a) $L^{-2}$ and (b) $\ln(L)L^{-2}$ scalings (see text).}
\end{table*}

In step 6a, the range of the potential was modified by varying $\lambda$
from $0$ to $0.894$. At $\lambda=1.0$, the range of the flat wall would
be zero.  Clearly, to chose an appropriate final value of $\lambda$
in step 6a, the strategy would be to first compute the accumulated free
energy difference up to step 5. Depending upon the desired accuracy, a
value is chosen which is much less than the statistical and numerical
errors in the combined steps 1-5.  From the ratio of this value and
$\Delta F_{\rm 1}/A$, a factor is obtained and multiplying it with the
range of the flat-wall potential in step 1, a new range can be calculated,
from which the appropriate final value of $\lambda$ in step 6a can be
deduced from Eq.~\ref{eq:flatgwall}.  By this strategy, step 6b would
become redundant as argued in Section~\ref{sec_ti}.

In Fig.~\ref{fig:step6TI}, we plot the thermodynamic integrands
corresponding to the forward and reverse paths of step 6a. Though the
curves are a bit noisy compared to the previous steps, the magnitude of
the integrands are also very small and hysteresis between the forward
and reverse calculations is $0.0005k_{\rm B}T$, while ${\Delta F_{\rm
6a}}/A=-0.003k_{\rm B}T$.  For the (110) and (111) orientations, the
contribution from step 6a is, $-0.020\pm0.001$ and $-0.002\pm0.001$,
in units of $k_{\rm B}T/\sigma^2$.

We have carried out several independent runs corresponding to step 6b and
from the runs with the least hysteresis, we extracted the free-energy
difference.  However, for all the three orientations the magnitude of
the contribution was less than $0.0005k_{\rm B}T/\sigma^2$ and hence
the thermodynamic integrands are not shown and neither were they added
to the final value of $\gamma_{\rm cl}$.

The final value of $\gamma_{\rm cl}$ after adding the contributions
from each step is reported in Table I for the three orientations and for
various system sizes corresponding to the (100) orientation. Using the
scaling parameter $\sigma^{\rm eff}= 1.0035\sigma$, the equivalent
interfacial free-energy for the hard-sphere potential is also
specified. Data from the cleaving-wall method corresponding to
the pure hard-sphere potential~\cite{davidchack2010} and that for
the inverse-power potential extrapolated to the hard-sphere limit
($n\rightarrow \infty$)~\cite{laird-davidchack2005prl} is also
reported. While our results are slightly higher than those obtained
for the former case, they are in good agreement with the later, within
the reported errors. Another TI approach reported recently yielded a
similar value for the (100) orientation of the hard-sphere crystal-liquid
interface~\cite{espinosa2014}.

To study systematic errors arising from finite-size effects we have
carried out simulations for the (100) orientation of the crystal-liquid
interface at various lateral dimensions ($L_{\rm x} \times L_{\rm y}$)
keeping the longitudinal dimension fixed at $78.58\,\sigma$. Recently,
Schmitz {\it et al.}~\cite{schmitz2014} identified the finite-size
corrections and proposed a scaling relation to obtain reliable estimates
for the interfacial free energies in the thermodynamic limit. According
to them, the leading finite size corrections to the interfacial
free energy in the thermodynamic limit is described by the following
relation~\cite{schmitz2014,binder82}:
\begin{equation}
\gamma_{\rm L,L_{\rm z}} =
\gamma_{\rm \infty} -A\frac{\ln L_{\rm z}}{L^2} + B\frac{\ln L}{L^{2}} 
+ \frac{C}{L^{2}}\,
\label{eq:finitesize}
\end{equation}
where $A$, $B$ and $B$ are constants and $L=L{\rm x}=L_{\rm y}$
corresponds to the lateral dimension of our system.

The second term in Eq.~(\ref{eq:finitesize}) is identified with the
translational entropy of the system arising from the movement of the
crystal-liquid interface. Since, the flat walls restrict the movement of
the crystal-liquid interface, this term can be neglected in our TI scheme.
In Fig.~\ref{fig:gam_vs_l2inv}, we plot $\gamma_{\rm cl}$ as a function
of $1/L^{\rm 2}$ and $\ln(L)/L^{\rm 2}$ separately.  Extrapolating the
data linearly yields $\gamma_{\rm cl}$, the equivalent hard-sphere
crystal-liquid interfacial free energy in the thermodynamic limit and
for the $1/L^{\rm 2}$ and $\ln(L)/L^{\rm 2}$ scalings we obtain the
values $0.599 k_{\rm B}T/{\sigma}^2$ and $0.601 k_{\rm B}T/{\sigma}^2$,
respectively.

In comparison to the values of $\gamma_{\rm cl}$ corresponding to the
smallest system sizes considered, the thermodynamic limit value is around
$3\%$ higher, indicating that finite-size corrections cannot be ignored
in order to obtain a reliable estimate.  The linear scaling observed in
Fig.~\ref{fig:gam_vs_l2inv} also indicates that capillary fluctuations are
not suppressed on account of the flat walls.  It is also to be noted that
values of $\gamma_{\rm cl}$ obtained in the thermodynamic limit for the
(100) orientation of the crystal-liquid interface are about $6-7\%$ less
than those corresponding to the capillary fluctuation~\cite{hartel2012}
and tethered Monte Carlo approach~\cite{martin2012}. More research is
needed to understand the origin of this discrepancy.

\section{Conclusion}
\label{sec_conc}
We have obtained the crystal-liquid interfacial free energy for the
hard sphere system via a novel thermodynamic integration scheme.  Good
agreement of our results with other computational approaches indicates
the success of our TI scheme even for short-ranged interaction potentials.
The flat Gaussian walls introduced in this scheme circumvents the problem
of achieving a reversible transformation due to the movement of the
crystal-liquid interface. These flat walls suppress the movement of this
interface and at the same time do not affect the capillary fluctuations.
Use of frozen-in crystalline layers to act as structured walls induces
ordering in the liquid compatible with the crystal structure and leads
to a smooth thermodynamic transformation to the desired final state.

Our results also indicate that finite-size errors have to be accounted
for in order to obtain accurate estimates for $\gamma_{\rm cl}$. The
accurate values for the interfacial free energy obtained in this work
can be used to validate future DFT approaches, which in the past have
yielded higher values as compared to a TI calculation~\cite{hartel2012}.
Furthermore, a TI calculation of $\gamma_{\rm cl}$ using our scheme is
necessary for the pure hard sphere system to be able to compare with
existing values. Work along these lines is currently under way.

{\bf Acknowledgments:}\\
The authors acknowledge financial support by the German DFG SPP 1296,
Grant No.~HO 2231/6-3.

\end{document}